\documentclass[doublecol]{epl2} 
\pdfoutput=1
\usepackage{subfigure}
\usepackage{amssymb}
\usepackage{amsfonts}
\usepackage{epsfig}
\usepackage{graphicx}
\usepackage{color}

\title{Universal fluctuations in radial growth models belonging to the KPZ universality class}
\shorttitle{Universal fluctuations in radial growth models} 

\author{Sidiney G. Alves\and Tiago J. Oliveira 
\and Silvio C. Ferreira 
\thanks{On leave at Departament de F\'{\i}sica i Enginyeria Nuclear, 
Universitat Polit\'ecnica de Catalunya, Barcelona, Spain.} }
\shortauthor{Alves \etal}

\institute{                    
  Departamento de F\'{\i}sica - Universidade Federal de Vi\c{c}osa, 36571-000, 
Vi\c{c}osa, Minas Gerais, Brazil
}

\abstract{
We investigate the radius distributions (RD) of surfaces obtained with
large-scale simulations of radial clusters that belong to the KPZ universality
class. For all investigated models, the RDs are given by the Tracy-Widom
distribution of the Gaussian unitary ensemble, in agreement with the conjecture
of the KPZ universality class for curved surfaces. The quantitative agreement
was also confirmed by two-point correlation functions asymptotically given by
the covariance of the Airy$_2$ process. Our simulation results fill the last
lacking gap of the conjecture that had been recently verified analytically and
experimentally. }

\begin{document}

\maketitle

Growth phenomena remain a topic of great interest in nonequilibrium Statistical
Physics, mainly because of the self-similarity and universality emerging from
dynamical local processes of different systems. In this context, one of the most
important examples is the Kardar-Parisi-Zhang (KPZ) universality class
introduced by equation~\cite{KPZ}: \begin{equation} \frac{\partial h}{\partial
t} = \nu \nabla^2 h+\lambda |\nabla h|^2+\eta, \end{equation} where $\eta$ is a
Gaussian noise. This universality class was observed in a large number of
models~\cite{barabasi,meakin} and a few experimental
systems~\cite{miettinen,TakeSano,TakeuchiSP}.

Former works on surface dynamics were mainly concerned with the scaling
properties of the surface fluctuations by means of scaling
exponents~\cite{barabasi,meakin,krug}. However, there is a number of other
universal quantities that are also suitable for determining universality in
{surfaces~\cite{TakeuchiSP}}. Examples include the stationary
distributions of the global interface width and the  extremal
height~\cite{tiago1,tiago2}, and the height distributions during the transient
regime~\cite{johansson,PraSpo1} that precedes saturation of the interface width.

The distributions during the transient (growth) regime were computed exactly for
some models in the KPZ class, in $1+1$
dimension~\cite{johansson,PraSpo1,krugrev,Ferrari}. Among the most relevant
cases, the height distribution (HD) of the single step model~\cite{barabasi},
determined analytically by Johansson~\cite{johansson}, has the Tracy-Widom
distribution of the largest eigenvalue of the Gaussian unitary ensemble (GUE) of
the random matrix theory~\cite{TW1} as a limit solution. Pr\"ahofer and Spohn
\cite{PraSpo1,PraSpo2} also obtained analytically the scaling form of the KPZ
universality class and a Tracy-Widom distribution for the HD of the polynuclear
growth model (PNG). Furthermore, they showed that the  HD for the growth from
flat substrates is given by the Gaussian orthogonal ensemble (GOE), while growth
from a single seed (radial growth) leads to a GUE distribution for the
radii~\cite{PraSpo1,PraSpo2}. Recently, Sasamoto and Spohn \cite{SasaSpo1} found
a solution of the one-dimensional KPZ equation with an initial condition that
induces the growth of curved surfaces  and the GUE distribution was confirmed as
the limit for the radius fluctuations. Subsequently, a numerical evaluation of
this solution corroborated these analytical results~\cite{Prolhac}. Pr\"ahofer
and Spohn~\cite{PraSpo3} also obtained an analytical solution for the limiting
process describing the surface fluctuations in the PNG model as the so-called
Airy$_2$ process.

Experimentally, the GOE and GUE distributions were obtained in a few experiments
exhibiting KPZ scaling. An evidence of the KPZ exponents was found in the slow
combustion of paper sheets~\cite{miettinen}. The burning fronts evolve from a
flat initial condition and the obtained HD has a reasonable agreement with GOE.
A recent experiment on electroconvection of turbulent liquid crystal
films~\cite{TakeSano} allowed to investigate an isotropic radial growth with
high accuracy. It was shown that this system belongs to KPZ universality class
exhibiting radius distribution (RD) in excellent agreement with GUE, including
the cumulants from second to fourth order. However, the mean is shifted and
tends to the GUE value as a power law $t^{-1/3}$. The same result was found by
Sasamoto and Spohn \cite{SasaSpo1} in a solution of the KPZ equation in $1+1$
dimensions with an initial edge condition indicating a universal behaviour of
the exponent 1/3. However, Ferrari and Frings~\cite{Ferrari} have
recently shown that the scaling law featuring this approach to GUE mean is not
universal. Indeed, they have shown that the mean shift decays as $t^{-1/3}$ for
the totally asymmetric simple exclusion process (ASEP) whereas no correction (up
order $\mathcal{O}(t^{-2/3})$) is found for the weakly  ASEP, both models
belonging to the KPZ class. The liquid crystal film setup was also used to
induce a front growth from a flat surface, and a good agreement with GOE was
obtained for the HDs~\cite{TakeuchiSP}.

The theoretical and experimental evidences above mentioned strongly suggest that
the GUE distribution is an universal feature of one-dimensional growth with
radial symmetry in the KPZ class. However, this conjecture is based on a limited
number of models with exact results~\cite{PraSpo1,PraSpo2,SasaSpo1,Ferrari} and
a single experimental work~\cite{TakeSano}. Numerical confirmation of
the GUE distributions are, up to this moment, missing. In order to fill this
gap,  we investigate the RD of large radial clusters (larger than $3\times 10^9$
particles) generated with different versions of the Eden model
\cite{eden,LetSil,SilSid1}.  We show that the RDs exhibit very good agreement
with GUE distribution. We also observed that the cumulants of the distribution
converge to the GUE values, as previously observed in other
systems~\cite{TakeSano,SasaSpo1}. The two-point correlation function also
converges to the Airy$_2$ process, as predicted by the
conjecture~\cite{PraSpo3}.

The Eden model~\cite{eden} consists in adding new particles in the empty
neighbourhood of a growing cluster. If the growth starts with a single particle,
the model yields asymptotically spherical clusters with a self-affine surface
exhibiting the scaling exponents of the KPZ universality class~\cite{SilSid1}.
We simulated off-lattice clusters in two-dimensions with the usual
algorithm~\cite{SilSid1,BJP}: a particle in the active (growing)
zone\footnote{The active zone consists of those particles having sufficient
empty space in its neighbourhood to add at least a new particle.} is selected at
random and a new particle is added in a random position chosen in the empty
neighbourhood of the selected particle. The procedure is repeated while the
cluster does not reach $N$ particles. With suitable
optimizations~\cite{SilSid1,BJP}, we were able to grow clusters with up to
$3\times 10^7$ particles. Since we randomly pick up a particle from a constantly
updated list containing $N_s$ surface sites, the time step is simply $\Delta
t=1/N_s$. A total number of up to $10^3$ off-lattice clusters were
grown in order to perform statistical averages.

We also have simulated Eden models on a square lattice using an algorithm
proposed in Ref. \cite{LetSil} that removes the lattice anisotropy effects. The
method consists in accepting a given growth step with probability
$p_j=(n_j/4)^\nu$, where $n_j=1,2,3,4$ is the number of occupied
nearest-neighbours (NN) of a selected growth site $j$ and $\nu$ is an adjustable
parameter. This method allows to generate isotropic clusters containing  more
than $4\times 10^{9}$ particles. We investigated two algorithms for lattice
simulations. In the version Eden A, one site is randomly selected
among all $N_g$ growth sites (empty NNs of the cluster) and then occupied.  In
the version Eden B, one of the $N_b$ sites in the cluster border and one of its
empty NN are randomly chosen and the empty one is occupied. The values of the
parameter $\nu$ that produce isotropic clusters are $\nu_A=1.72$ and $\nu_B=1$
for Eden A and B, respectively~\cite{LetSil}.  At each attempt, the time is
increased by $1/N_g$ and $1/N_b$ for Eden A and B, respectively,  independently
of the growth success. The numbers of clusters used for statistics were up to $10^4$
for Eden A and up to $10^3$ for Eden B. Notice that statistical fluctuations are
much stronger in Eden A due to the higher amount of overhangs in the surface.

\begin{figure}[t]
\centering
\includegraphics[width=7.5cm]{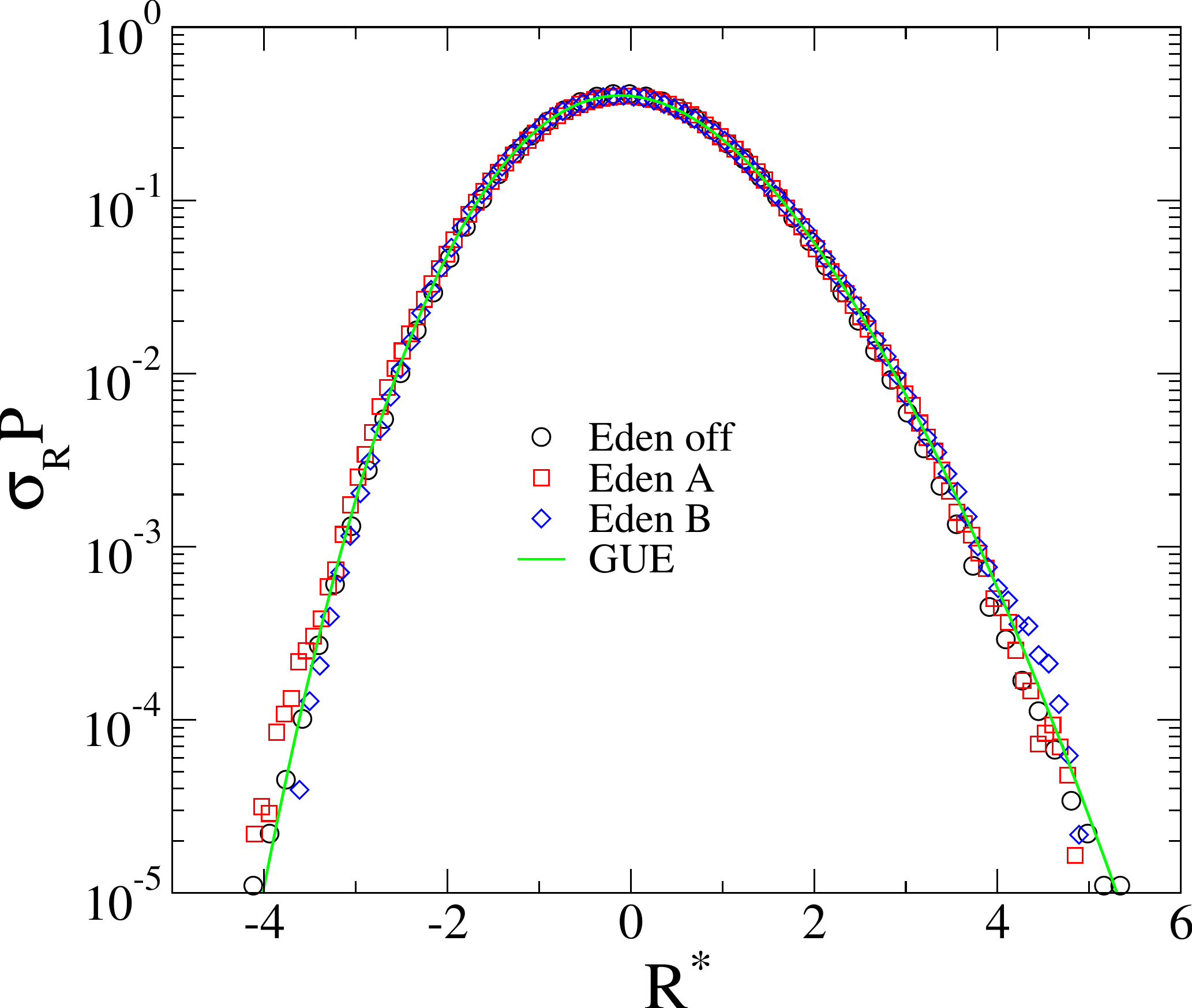}
\caption{(Color online) Radius distributions of on- and off-lattice Eden models
rescaled to a null mean  and a unitary variance. The mean radius of the
aggregates are approximately $2500$ for the off-lattice
model and $3.2 \times 10^{4}$ for lattice models. The solid line is the rescaled
GUE distribution. In this plot, $R^*\equiv(R-\langle R\rangle)/\sigma_{R}$.}
\label{dist_tiago}
\end{figure}

\begin{figure*}[ht]
\centering
\subfigure[]{\includegraphics[width=5.4cm]{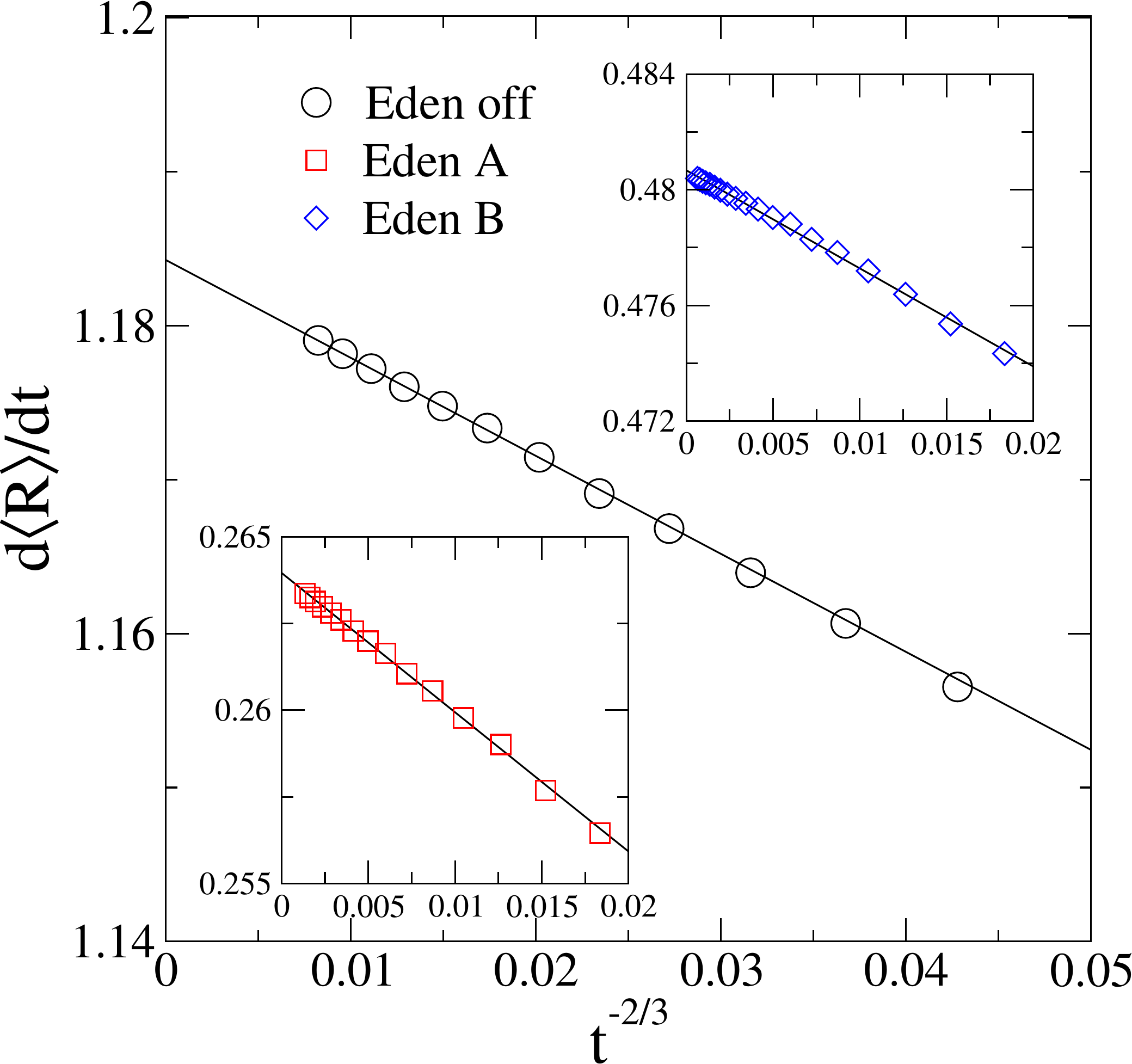}} ~ 
\subfigure[]{\includegraphics[width=5.5cm]{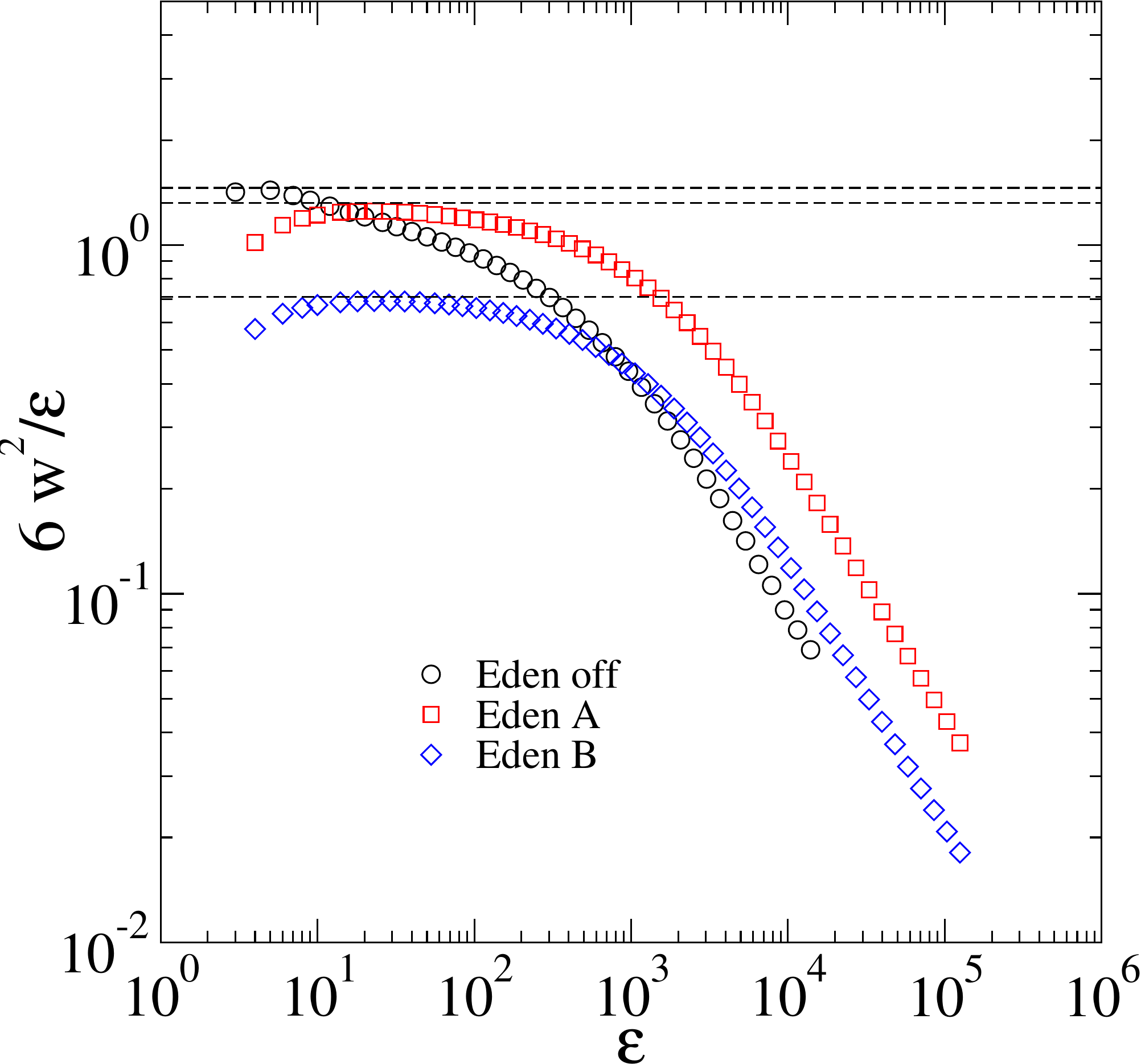}} ~ 
\subfigure[]{\includegraphics[width=5.5cm]{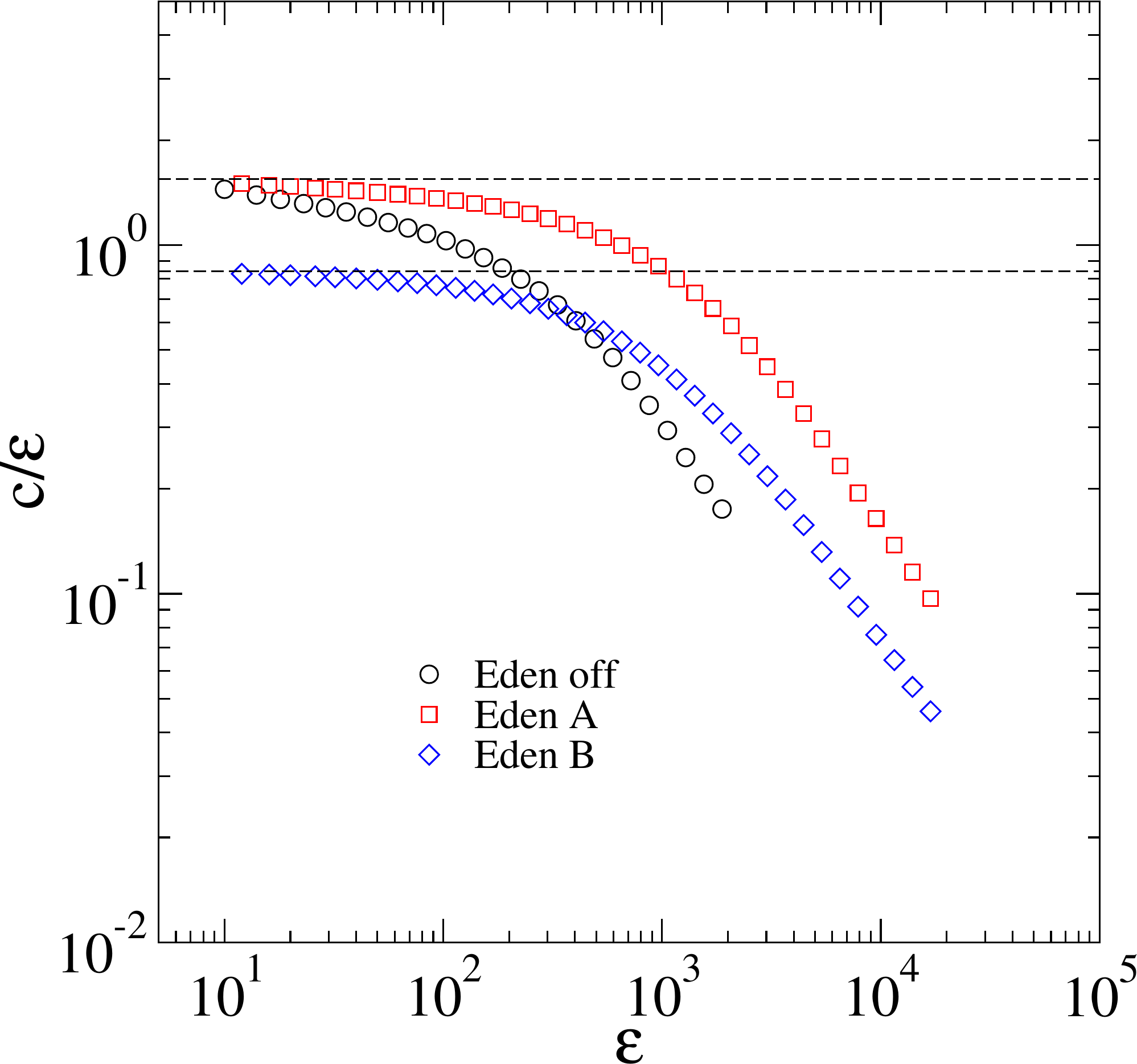}} ~ 
\caption{(Color online) Determination of the parameters related to radius
evolution given by Eq. (\ref{eqPS}). (a) Average radius growth rate
$d\langle{R}\rangle/dt$ against $t^{-2/3}$ for off- (main plot) and on-lattice
(insets) Eden models. Dashed lines are extrapolations to $t\rightarrow\infty$
used to determine $\lambda$. (b) Rescaled  interface width for a scale
$\epsilon$. The horizontal lines are estimates of the parameter $A$. Cluster
average radii are 32000 for on-lattice and 2700 for off-lattice models. (c) The
equivalent analysis of (b) for the correlation function.}
\label{hwl}
\end{figure*}

We start the analysis with a comparison among the RD of the Eden models and
the GUE distribution, both suitably rescaled and shifted to have a null mean and
a unitary variance. We assume a scaling form 
\begin{equation}
 P(R) = \frac{1}{\sigma_{R}} G \left( \frac{R- \langle R\rangle }{\sigma_{R}} \right), 
\label{eqRET}
\end{equation}
where $\sigma_{R}^2$ is the variance of the RD and $G(x)$ is a normalized
scaling function. This scaling form reduces finite size corrections, what has
improved data collapses in other analyses, as interface width
distributions~\cite{tiago1}, for example. In Fig.~\ref{dist_tiago}, we compare
the rescaled RDs for the three Eden models with the rescaled GUE distribution. An
excellent collapse of all curves upon a single curve $G(x)$ was obtained.
Similar results hold for different growth times. This data collapse confirms
that, a part of corrections to scaling in the cumulants described below, the RDs
of the all investigated models agree with the GUE distribution, as conjectured
by Pr\"ahofer and Spohn~\cite{PraSpo1}.

In radial growth belonging to the KPZ universality class, the radii are
stochastic variables evolving in time as~\cite{PraSpo1,PraSpo2}:
\begin{equation}
 R(t) \simeq \lambda t + {\left( A^{2} \lambda t/2 \right) } ^{1/3} \chi_{GUE},
\label{eqPS}
\end{equation}
where $\lambda$ and $A$ are two non-universal (model dependent) parameters. The
random variable $\chi_{GUE}$ is  distributed according to the GUE Tracy-Widom
distribution~\cite{TW1}. Therefore,  $\lambda$ is the asymptotic radial growth
rate obtained from $\lambda \simeq d\langle{R}\rangle/dt + a_{v} t^{-2/3}$ in the
limit $t \rightarrow \infty$~\cite{krug1}. In Fig. \ref{hwl}(a), we show the
average radius growth rate against a power of time.  The extrapolated
asymptotic values are $\lambda \approx 1.1843(2)$ for off-lattice, $\lambda
\simeq 0.2639(2)$ for Eden A, and $\lambda\simeq 0.4807(2)$ for Eden B, where
the uncertainties obtained in the regressions are shown in parenthesis.

The parameter $A$ was estimated in two independent ways. We can use the local
squared surface roughness, in a window of size $\epsilon$, defined as 
\begin{equation}
 w^{2}(\epsilon,t) =\langle [R(x,t)]^2 \rangle_{\epsilon} -\langle R(x,t) \rangle_{\epsilon}^{2} ,
\end{equation}
where $\left\langle\dots \right\rangle_\epsilon$  denotes an average within
several windows in the interface.
Alternatively, we can estimate the $A$ value using the height-height correlation
function 
\begin{equation}
c(\epsilon,t) = \langle [ R(x+\epsilon,t) - R(x,t)]^{2}\rangle_\epsilon. 
\label{eqcor}
\end{equation}
For long times, theoretical arguments predict that $w^{2}\simeq A \epsilon /6$
and $c \simeq A \epsilon$~\cite{krug1}. Curves for $6 w^{2}/\epsilon$ as
functions of $\epsilon$ are shown in Fig. \ref{hwl}(b). Well-defined plateaus
are observed for both on-lattice models, except for  short scales, when a small
deviation is observed.  Since off-lattice simulations are much smaller, the
plateau is not so evident as in the on-lattice case, but the data also tend to a
constant value. For sake of comparison,  we measured the local roughness
exponent, defined as $w(\epsilon)\sim \epsilon^\alpha$, and found $\alpha\approx
0.43$ for our largest off-lattice simulations. This value is considerably
smaller than the expected exponent of the KPZ class $\alpha=1/2$, confirming the
presence of strong finite time effects in the exponents. Roughness exponent for
the on-lattice models are very close to the value $\alpha=1/2$. The estimates of
parameter $A$ are represented by dashed lines in Figs. \ref{hwl}(b) and (c). The
estimates using local interfaces width are: $A\approx1.46$ for off-lattice,
$A\approx1.32$ for Eden A, and $A\approx0.71$ for Eden B models. The correlation
function estimates are slightly larger: $A\approx1.55$ for off-lattice and A
models, and $A\approx0.84$ for Eden B.

\begin{figure}[hbt]
\centering
\includegraphics[width=6.6cm]{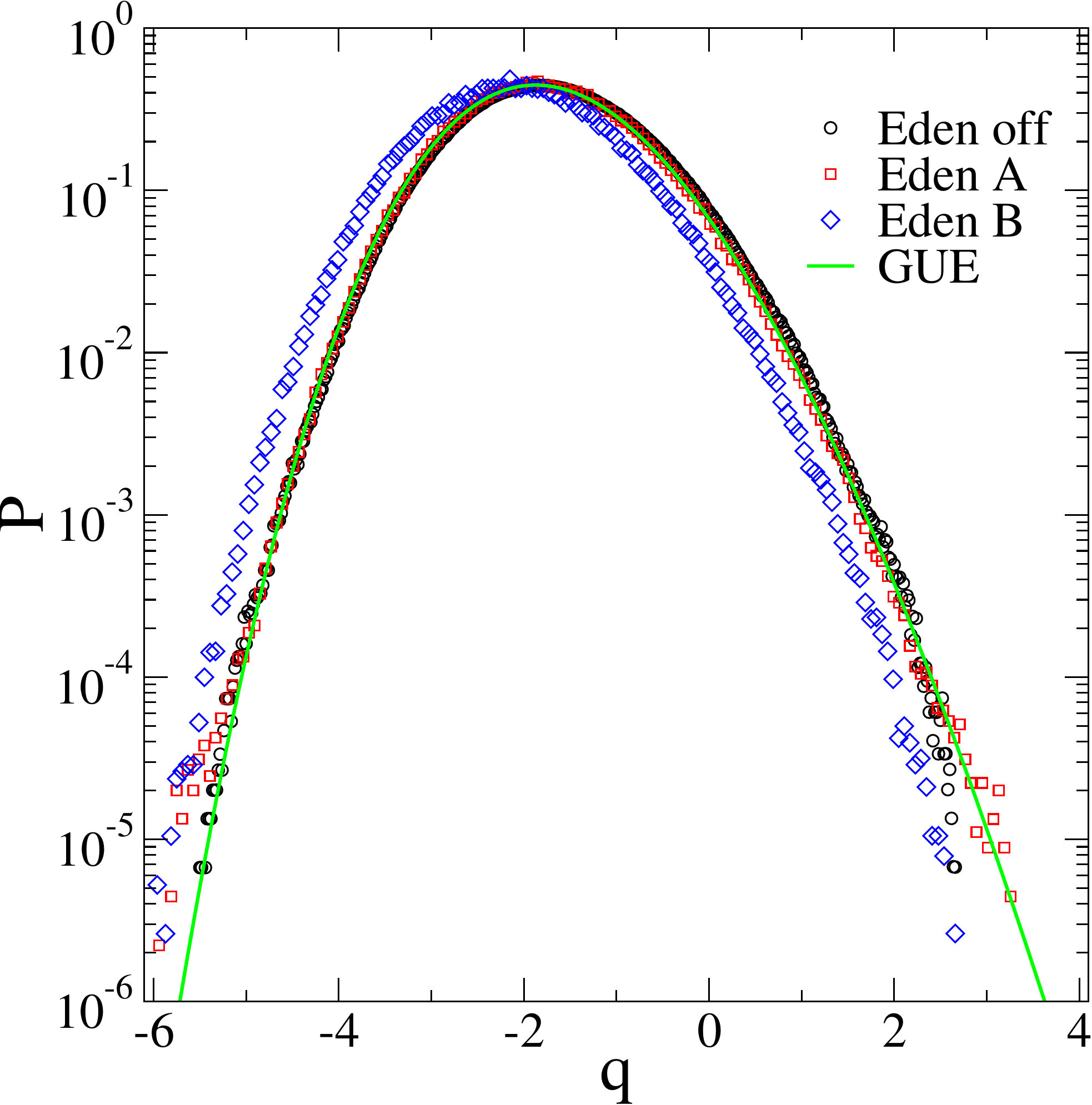}
\caption{{(Color online) Probability distribution of $q = (R - \lambda
t)/(A^2 \lambda t/2)^{1/3}$ for our largest-time simulations: Off-lattice Eden
model with cluster mean radius $\bar{r}=2700$; Eden  A and B models with mean radius 
$\bar{r} = 32000$. The solid line is the GUE distribution.}} 
\label{eden_gue}
\end{figure}

\begin{figure}[ht]
\centering
\subfigure[]{\includegraphics[width=5.5cm]{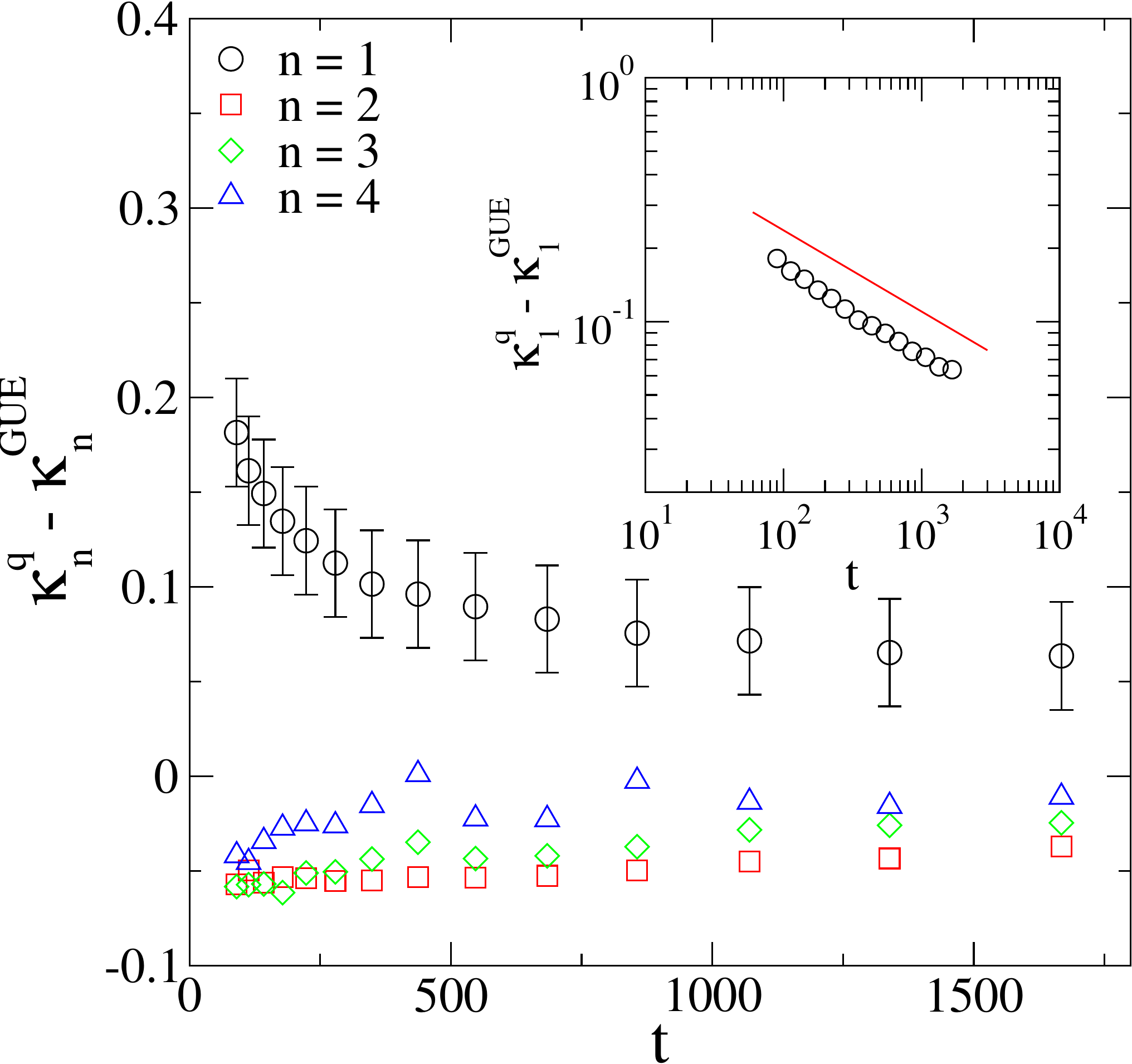}} 
\subfigure[]{\includegraphics[width=5.5cm]{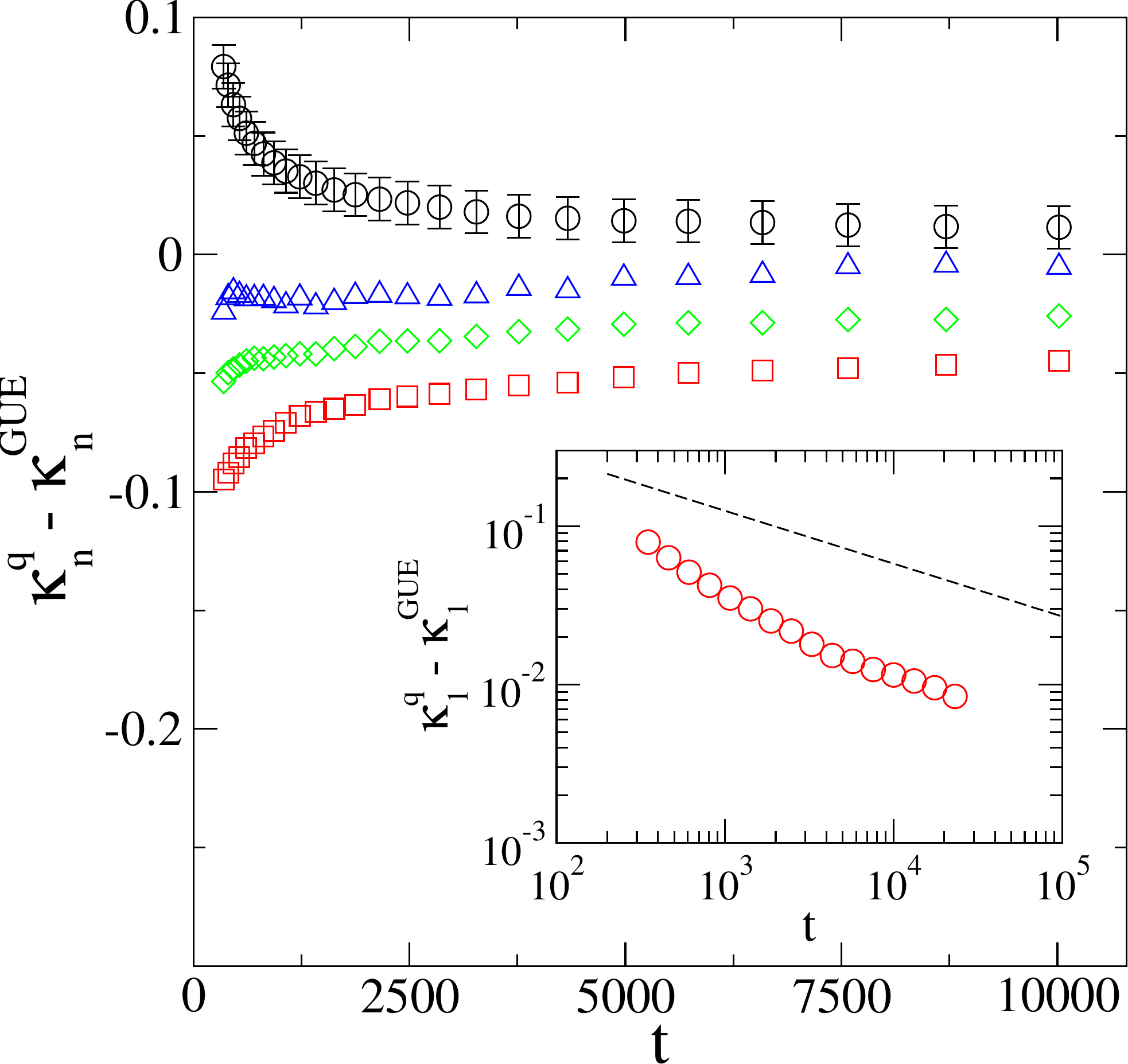}} 
\subfigure[]{\includegraphics[width=5.5cm]{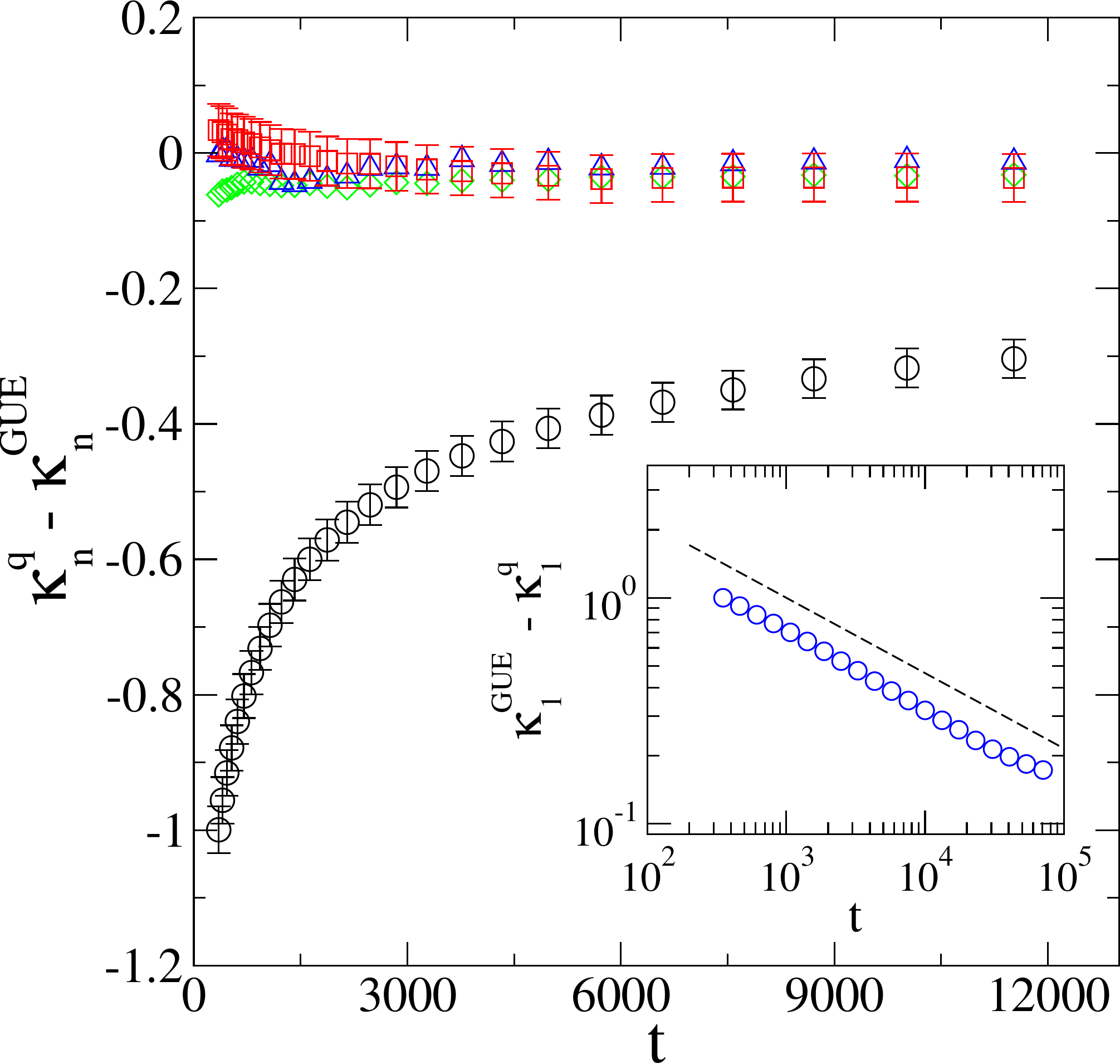}} 
\caption{(Color online) Differences between cumulants obtained for
Eden models and GUE distribution as functions of time for (a) off-lattice
simulations, (b) Eden A, and (c) Eden B. Cumulants up to fourth order are shown
in the main plots. Insets show the mean against time in a double 
logarithmic scale. Dashed lines have slope -1/3.} 
\label{cumul}
\end{figure}

In agreement to Eq. (\ref{eqPS}), the quantity 
\begin{equation}
 q = \frac{R - \lambda t}{(A^2 \lambda t/2)^{1/3}}
 \label{eq_q}
\end{equation}
is a  random variable given by a GUE distribution. The RDs shown in
Fig.~\ref{eden_gue} were obtained with the values $\lambda = 1.1842$ and
$A=1.45$ for off-lattice simulations, $\lambda = 0.263887$ and $A=1.43$ for Eden
A, and $\lambda = 0.4806$ and $A=0.805$ for Eden
B. 
As predicted by radial KPZ conjecture, a very well
collapse for different models (and different times) is observed. The
agreement with the GUE distribution is noticeable, except by a shift
in $q$, that vanishes as $t\rightarrow\infty$. The shift is more evident for
Eden B as can also be seen in Fig.~\ref{cumul}. Our results confirm the
limiting scaling form conjectured by Ph\"ahoffer and Spohn~\cite{PraSpo1,PraSpo2}, as
previously observed in theoretical~\cite{krugrev,PraSpo1,PraSpo2,SasaSpo1} and
experimental~\cite{TakeSano} systems.

In order to quantify the agreement between RDs and GUE distributions, we
investigate the $n$th order cumulants of the probability distribution $P(q)$
denoted by $\kappa_n^q$. The differences between the cumulants obtained for
off-lattice simulations and the GUE values are shown, as function of time, in
Fig. \ref{cumul}(a). As expected, all cumulants converge to the GUE values. As
observed experimentally by Takeuchi and Sano~\cite{TakeSano}, the first moment
decreases towards the GUE value while the higher order cumulants increases
towards GUE values. The same happens for Eden A,  as can be seen in
Fig.~\ref{cumul}(b). Differently, the mean for Eden B increases (more slowly
than the others) towards GUE and the higher order cumulants converges more
quickly to the theoretical values. This negative amplitude of the difference
between simulation and GUE mean was also observed in a solution of the KPZ
equation~\cite{SasaSpo1}. Indeed, the law describing the convergence of the mean
have recently attracted great interest~\cite{TakeSano,SasaSpo1,Ferrari}. Our off-lattice
and Eden B simulations are very well described by the $t^{-1/3}$ law 
previously reported~\cite{TakeSano,SasaSpo1} while the simulation of Eden A are
only consistent with this approach since a long power law regime was not
observed.

\begin{figure}[t]
\centering
\includegraphics[width=7.0cm]{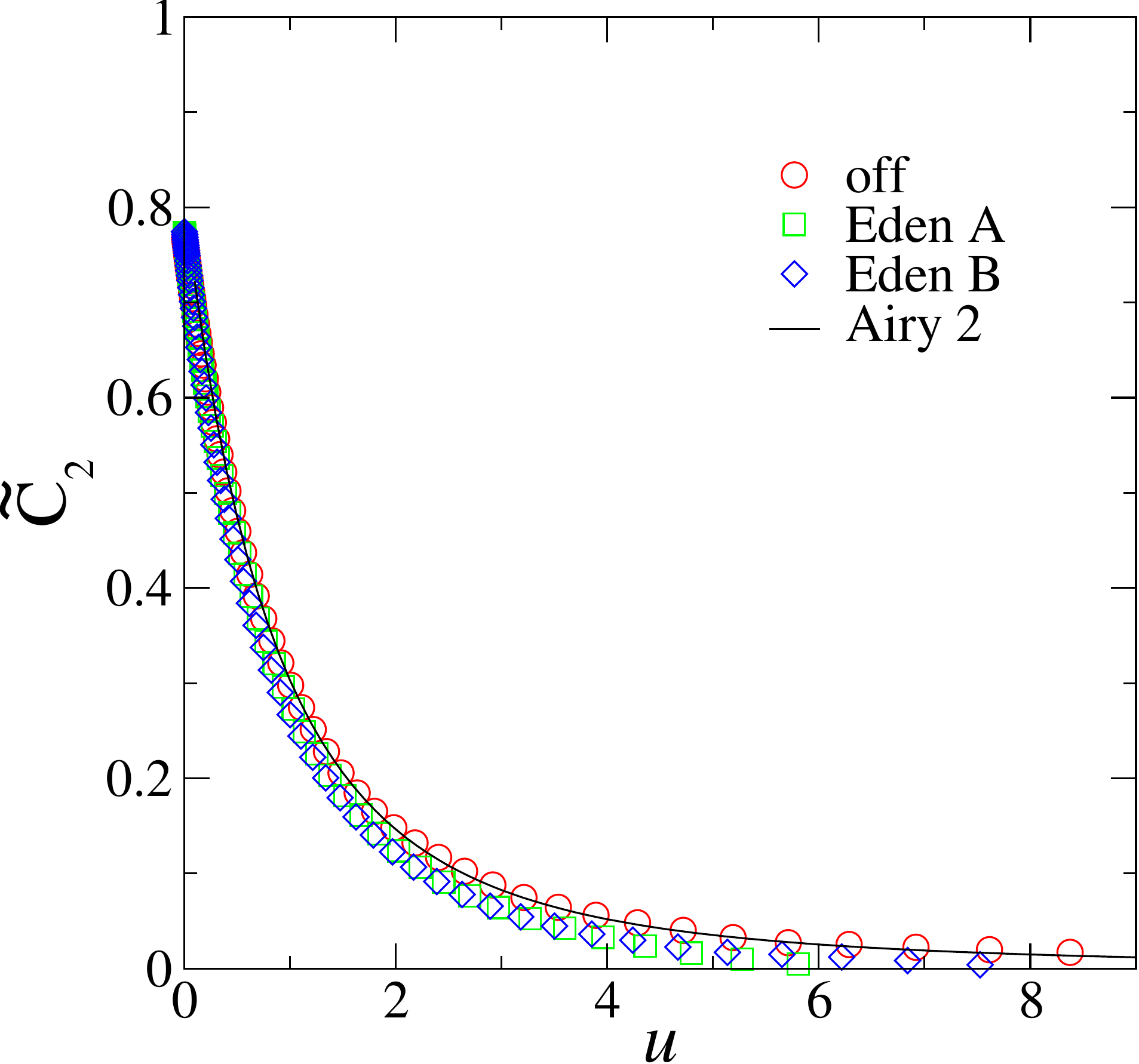}
\caption{(Color online) Scaled two-point correlation function for the Eden
models. The average radius of the clusters are 2700 for off-lattice and 32000 for
A and B models, respectively. The solid curve is the covariance of
the Airy$_2$ process. In this plot, $\widetilde{C}_2 = (A^2 \lambda t/2)^{-2/3}C_2 $
and $u=(A \epsilon/2)(A^2 \lambda t/2)^{-2/3}$.}
\label{covar}
\end{figure}

A further evidence of the agreement between Eden growth and radial KPZ
conjecture is yielded by the two-point correlation function given by $C_{2}(\epsilon,t)
= \langle R(x+\epsilon,t)R(x,t)\rangle-\langle R\rangle^2$, that, in agreement with the
conjecture~\cite{PraSpo3}, scales at long times as $C_2(\epsilon,t) \simeq (A^2 \lambda t/2)^{2/3}
g_{2}(u)$ with $u = (A \epsilon/2)/(A^2 \lambda t/2)^{2/3}$, where
$g_{2}(u)$ is the covariance of the Airy$_{2}$
process~\cite{Bornemann}. In Fig. \ref{covar}, we show the correlation function
$C_2$ for different models. A very good agreement  between the scaled $C_{2}$ 
for all Eden models and $g_{2}$ function is obtained, showing that the models
are well described by the Airy$_2$ process. It is worth mentioning that scaled
$C_2$ approaches $g_2$ for long times and that this approach is slower for 
on-lattice models.

In summary, the KPZ universality class in radial growth has been subject of
recent analytical~\cite{SasaSpo1} and experimental~\cite{TakeSano}
investigations that agreed with the conjecture proposed by Pr\"ahofer and Spohn
\cite{PraSpo1,PraSpo2}, where interface fluctuations in systems belonging to the
KPZ universality class are described by well-known universal distributions.
However, a computational verification in growth models was lacking until the
present work. In the present Letter, we have investigated the radius
distributions in clusters obtained with Eden growth models~\cite{SilSid1,LetSil}
starting from a single particle. 

The radius distributions obtained for all models exhibit an excellent
agreement with the scaling ansatz given by Eq.~(\ref{eqPS}), that associates the
radius fluctuations with the Tracy-Widom~\cite{TW1} distribution of the Gaussian
Unitary Ensemble. The cumulants of order $n\ge 1$ associated to RD converge to
the corresponding GUE cumulants  for all investigated models. A finite time
correction of order $t^{-1/3}$ in the first moment was clearly observed for
off-lattice and Eden B simulations, in agreement with other
systems~\cite{TakeSano,SasaSpo1}. Finally, a correlation function in accordance
with the so-called Airy$_2$ process yields a further strong evidence that the
radius fluctuations in all investigated growth models are in agreement with the
KPZ conjecture.

As a final remark, notice that the the small exponent $\alpha \approx
0.43$ obtained for the largest off-lattice simulations shows that scaling
exponents undergo strong finite time effects. Therefore, the RD analysis may be
more reliable than scaling exponents to determine the universality class of a
system. In particular, the scaling form given by Eq.~(\ref{eqRET}) is
simpler than the analysis with Eq.~(\ref{eqPS}), since fit procedures are not
required in the first approach and it does not have finite time corrections.

In conclusion, the  Pr\"ahofer-Spohn conjecture is now fully verified in the
three general branches of Statistical Physics: experimental, theoretical, and
computational.

\acknowledgments
 This work was partially supported by the Brazilian agencies CNPq, FAPEMIG, and
CAPES.  Authors thank F. Bornemann by kindly providing the covariance of the
Airy$_2$ process. We also thank the former discussions with Herbert Spohn that motivated
the beginning of this work. SCF thanks the kind hospitality at the Departament
de F\'{\i}sica i Enginyeria Nuclear/UPC.


\end{document}